# Hikami-Larkin-Nagaoka (HLN) treatment of the Magneto Conductivity of $Bi_2Te_3$ Topological Insulator

Rabia Sultana[1,2], Ganesh Gurjar[3], P. Neha[3], S. Patnaik[3] and V.P.S. Awana[1,2*]

[1] National Physical Laboratory (CSIR), Dr. K. S. Krishnan Road, New Delhi-110012, India

[2] Academy of Scientific and Innovative Research (AcSIR), NPL, New Delhi-110012, India

[3] School of Physical Sciences, Jawaharlal Nehru University, New Delhi-110067, India

We report the magneto-conductivity analysis at different temperatures under magnetic field of up to 5Tesla of a well characterized $Bi_2Te_3$ crystal. Details of crystal growth and various physical properties including high linear magneto resistance are already reported by some of us. To elaborate upon the transport properties of $Bi_2Te_3$ crystal, the magneto conductivity is fitted to the known HLN (Hikami Larkin Nagaoka) equation and it is found that the conduction mechanism is dominated by both surface driven WAL (weak anti localization) and the bulk WL states. The value of HLN equation coefficient (α) signifying the type of localization (WL, WAL or both WL and WAL) falls within the range of -0.5 to -1.5. In our case, the low field (±0.25Tesla) fitting of studied crystal exhibited value of α close to -0.86 for studied temperatures of up to 50K, indicating both WAL and WL contributions. The phase coherence length ($l_\varphi$) is found to decrease from 98.266 to 40.314nm with increasing temperature. Summarily, the short letter reports the fact that bulk $Bi_2Te_3$ follows the HLN equation and quantitative analysis of the same facilitates to know the quality of studied crystal in terms of WAL to WL contributions and thus the surface to bulk conduction ratio.

*Corresponding Author
Dr. V. P. S. Awana:  E-mail: awana@nplindia.org
Ph. +91-11-45609357, Fax-+91-11-45609310
Homepage: awanavps.webs.com

**Introduction**

Topological insulators (TIs) are the so called wonder materials of recent times. The TIs are known to be conventional insulator in their bulk and as a conductor at the edges/surface having gapless states, which are further protected by time reversal symmetry (TRS) [1-11]. Apart from their unique physical properties realized so far, TIs do act as challenging materials in condensed mater physics community owing to their unusual magnetic behaviour, which could possibly be used to find a variety of exotic physical phenomenon resulting into novel applications. As reported, intrinsic TIs exhibits two different type of magneto-resistance (MR) behaviour depending upon the applied magnetic field [12]. One of them is the WAL behaviour which is observed as a typical v type cusp (sharp MR dip) at lower magnetic field, whereas the other is the linear non saturating MR behaviour observed at higher magnetic fields. The WAL effect exhibits negative magneto conductivity behaviour, whereas the WL effect exhibits positive magneto conductivity behaviour at lower magnetic field and temperatures. However, intrinsic TIs ($Bi_2Te_3$, $Bi_2Se_3$ and $Sb_2Te_3$) exhibits WAL behaviour as long as the surface state gap remains unopened, but experiences a competing effect of both WAL and WL and a crossover from WAL to WL effect as the TRS is broken due to opening of a surface energy from say doping of magnetic impurities [13]. Moreover, the WAL behaviour is significantly affected depending upon the type of material such as thin film, bulk single crystals, nano - flakes and nano - wires due to size dependent interactions between the surface and bulk states or the electron – electron interactions leading to competing WAL and WL [14-16]. It is also known that WAL behaviour is responsive only to the perpendicular component of the applied magnetic field, which is further described by the HLN (Hikami Larkin Nagoka) model [17, 18]. The HLN model in fact nicely represents the surface versus bulk conduction contributions to the overall conductivity of the TIs. The two important parameters being considered in HLN model are the pre factor ($\alpha$) and phase coherence length ($l_\varphi$). Principally, the pre factor ($\alpha$) moves from -0.5 to -1.5, depending upon the contributions from WL and WAL or the competing conduction channels. Henceforth the fitting of magneto-conductivity of TIs to HLN model is very fruitful to know the resultant conduction process. Additionally, the phase coherence length ($l_\varphi$) in the HLN equation is found to be temperature dependent and exhibits a power-law behaviour as confirmed theoretically i.e. $l_\varphi \sim T^{-0.5}$ for 2D systems and $l_\varphi \sim T^{-0.75}$ for 3D systems [17]. Higher the temperature, lower the coherence length ($l_\varphi$). Recently, some of us reported detailed crystal growth and physical property characterization of one of the popular TI i.e.,

Bi$_2$Te$_3$ [19, 20]. Keeping in view the importance of the overall conduction process of a TI in terms of competing WAL and WL, in current short article, we report the HLN treatment of the magneto conductivity of our well characterized [19,20] Bi$_2$Te$_3$ single crystal.

**Experimental details**

Bulk single crystals of Bi$_2$Te$_3$ were grown by the self flux method via the conventional solid state reaction route. The detailed crystal growth mechanism is illustrated in ref. [19, 20]. In brief, stoichiometric ratio of Bi and Te powders were mixed thoroughly inside a glove box with Ar atmosphere. The mixed powder was pressed into a rectangular pellet, sealed in an evacuated quartz tube and was kept inside an automated programmable box furnace. Heated to 950°C for 7.5 hours, hold for 12 hours and then slowly cooled (2°C/hour) to 650°C followed by switching off the furnace. The obtained crystal was then taken out of the quartz tube and mechanically cleaved for further structural and magneto transport measurements. The phase identification of the resultant bulk single crystal of Bi$_2$Te$_3$ was carried out using Rigaku Miniflex II, Powder X-ray Diffractometer (PXRD) with Cu-Kα radiation (λ=1.5418 Å). The magneto transport measurements were done using a 14Tesla down to 2K Quantum Design Physical Property Measurement System (PPMS), Model 6000.

**Results and Discussion**

Figure 1 depicts the single crystal XRD pattern of the resultant Bi$_2$Te$_3$ crystal in the angular range of $2\theta_{min} = 10^0$ and $2\theta_{max} = 80^0$. The XRD pattern shows well indexed sharp (00l) reflections, indicating the good crystalline nature of the synthesized Bi$_2$Te$_3$ crystal. The inset (a) of Fig. 1 displays the temperature dependent electrical resistivity plots of as synthesized Bi$_2$Te$_3$ single crystal under different applied magnetic fields i.e., $\rho$(T)H. The temperature varies from 5K to 50K, whereas the applied magnetic field ranges from 0Tesla to 6Tesla. Here, the resistivity curves appear to increase with temperature, clearly indicating that the as synthesized Bi$_2$Te$_3$ single crystal exhibits a metallic nature. Further the $\rho$(T)H clearly shows that the resistivity increases substantially with applied field at particular temperature. The other inset of Fig.1 i.e., inset 1(b) shows the percentage change of MR under different applied magnetic fields and temperatures for the studied Bi$_2$Te$_3$ single crystal. The applied magnetic field is varied from 0Tesla to 5Tesla and the temperature ranges from 2.5K to 280K.The MR (%) is obtained using the formula MR (%) = {[ρ(H) - ρ(0)] /

ρ(0)}*100, where ρ(0) and ρ(H) represents the resistivity values under zero and non zero applied magnetic fields (H) respectively. At lower magnetic fields say below 3Tesla the MR curves at 2.5, 5 and 10K seems to overlap, but bifurcates as the field is increased say above 3Tesla. Also, the shape of the MR curve seems to exhibit a v-type shape at the lower temperature (2.5K) which gradually tends to broaden as the temperature is increased (5, 10 and 50K) and finally changes into a straight line shape with least MR at the highest temperature (280K). The calculated MR value for the lowest (2.5K) and highest (280K) temperatures appears to be about 250% and 5% respectively, under 5Tesla applied magnetic field. Consequently, the MR value is seen to decrease from 250% to about 5% with increase in temperature from 2.5K to 280K. Accordingly, we can say that the as synthesized $Bi_2Te_3$ single crystal exhibits a temperature dependent MR value under applied magnetic fields.

To study the transport properties more elaborately, we have fitted the low field magneto- conductivity curves of the bulk $Bi_2Te_3$ single crystals using the Hikami - Larkin - Nagaoka (HLN) as below; [18]

$$\Delta\sigma(H) = \sigma(H) - \sigma(0) = -\frac{\alpha e^2}{\pi h}\left[\ln(\frac{B_\varphi}{H}) - \Psi\left(\frac{1}{2} + \frac{B_\varphi}{H}\right)\right]$$

Where, $\Delta\sigma(H)$ represents change of magneto-conductivity, $\Psi$ is the digamma function, e is the electronic charge, h is the Planck's constant, $B_\varphi = \frac{h}{8e\pi H l_\varphi}$ is the characteristic magnetic field, H is the applied magnetic field, $l_\varphi$ is the phase coherence length and α is a coefficient indicating the type of localization (WL, WAL or both WL and WAL). The pre factor, α exhibits values depending upon the type of spin orbit interaction (SOI) and magnetic scattering [18]. Accordingly, α = 0 when the magnetic scattering is strong (unitary case), α = 1 when the SOI and magnetic scattering is weak or absent (orthogonal case) and α = -0.5 when SOI is strong and there is no magnetic scattering [18].

The coefficient α, determining the type of localization as well as the number of coherent transport channels should have values -0.5 for single surface conducting channel and between -0.5 to -1.5 for multi parallel conduction channels (surface and bulk states) [21-27]. However, the experimentally fitted value of α varies widely, due to the problems arising from differentiating the bulk and surface contributions clearly. As reported, α may lie between –0.4 and –1.1, for single surface state, two surface states, or intermixing between the surface and bulk states [23, 26].

Figure 2 shows the fitted magneto-conductivity curves of bulk $Bi_2Te_3$ single crystal and using HLN equation at different temperatures (2.5, 5, 10, and 50K) under applied magnetic fields of ± 2Tesla. Figure 2 clearly shows that at lower fields i.e., up to ± 2Tesla the magneto-conductivity curves at 2.5K, 5K and 10K seems to overlap on each other but bifurcates at higher fields and follows HLN behavior. However, to study the HLN equation more precisely and to extract the fitting parameters i.e., pre factor (α) and phase coherence length ($l_\varphi$) one need to fit the magneto-conductivity data in much lower magnetic fields i.e., where v type shape is seen in MR. For this very reason in Figure 3, we show the HLN fitted magneto conductivity data of studied $Bi_2Te_3$ single crystal at much lower applied magnetic fields of up to ± 0.25Tesla. The HLN fitted lines are indicated by solid lines and the fitting parameters (α and $l_\varphi$) are given in the Figure itself. Both α and $l_\varphi$ exhibit close values of around -0.85 and 95nm respectively at lower temperatures i.e., at 2.5, 5, and 10K. At higher temperature i.e., at 50K though the α value is close to lower temperatures the phase coherence length ($l_\varphi$) is decreased to nearly half (40nm).

The fitted values of pre factor (α) and phase coherence length ($l_\varphi$) at all the HLN fitted temperatures for $Bi_2Te_3$ crystal are given in Table 1. It is clear from Table 1 that though the value of pre factor α is nearly unchanged and remains within range of -0.8543 to -0.88, the coherence length ($l_\varphi$) decrease from 92.26 to 40.31nm at 2.5K and 50K respectively. At intermediate temperatures, of 5 and 10K, the α and $l_\varphi$ values are -0.85, -0.86 and 96.04 and 92.97nm respectively. The resultant value of α at studied temperatures i.e., at 2.5, 5, 10 and 50K of around -0.8543 to -0.88 demonstrates that the conduction mechanism is dominated by both WAL and WL as being originated from both the 2D surface and 3D bulk states of the studied TI [27, 28]. Further, at 50K the relatively lower value of coherence length (40.314nm) indicates the more dominance of WL over WAL. It is clear that our quality bulk $Bi_2Te_3$ crystal follow the HLN equation. The HLN analysis could be a fruitful exercise to know the quality of the topological insulators in terms of surface to bulk conduction ratio.

Summarily, we analysed the magneto-conductivity data of our quality $Bi_2Te_3$ bulk single crystal in terms of HLN equation and found that both the WAL and WL contributes to the conduction process. Further, it is seen that phase coherence length decreases to nearly half from around 98 to 40nm as the temperature increased from 2.5 to 50K.

**Acknowledgements**

The authors from CSIR-NPL would like to thank their Director NPL, India, for his keen interest in the present work. S. Patnaik thanks DST-SERB project (EMR/2016/003998) for the low temperature high magnetic facility at JNU, New Delhi. Rabia Sultana and Ganesh Gurjar for CSIR research fellowship. P. Neha thanks UGC for BSR fellowship. Rabia Sultana also thanks AcSIR-NPL for Ph.D. registration.

**Figure Captions**

**Figure 1:** X-ray diffraction pattern of as synthesized $Bi_2Te_3$ single crystal. Inset **(a)** temperature dependent electrical resistivity of $Bi_2Te_3$ single crystal in temperature range of 5 to 50K and magnetic field varying from 0Tesla to 6Tesla **(b)** MR (%) as a function of magnetic field (H) perpendicular to ab plane at different temperatures for $Bi_2Te_3$ single crystal.

**Figure 2:** WAL related magneto-conductivity for $Bi_2Te_3$ single crystal at different temperatures (2.5K, 5K, 10K and 50K), fitted using the HLN equation up to ± 2Tesla.

**Figure 3:** Magneto-conductivity curves for $Bi_2Te_3$ single crystal at different temperatures (2.5K, 5K, 10K and 50K), fitted using the HLN equation up to ± 0.25Tesla

**Table 1** HLN fit values of pre factor (α) and phase coherence length ($l_\varphi$) for $Bi_2Te_3$ crystal

| Temperature | α | l$\phi$ |
|:---:|:---:|:---:|
| 2.5K | -0.854 | 98.266 nm |
| 5K | -0.855 | 96.045 nm |
| 10K | -0.86 | 92.979 nm |
| 50K | -0.88 | 40.314 nm |

**Fig.1**

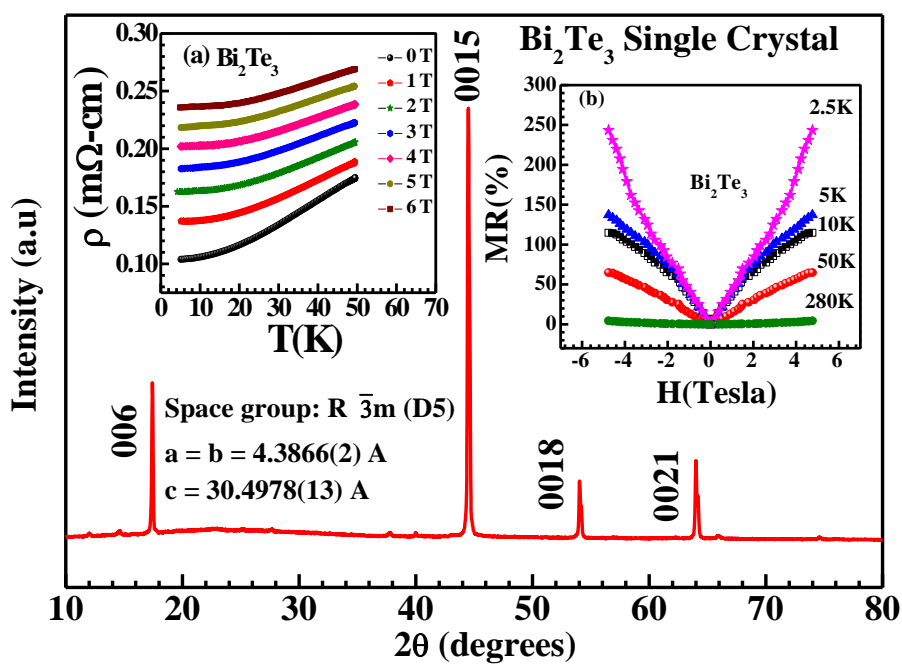

**Fig. 2**

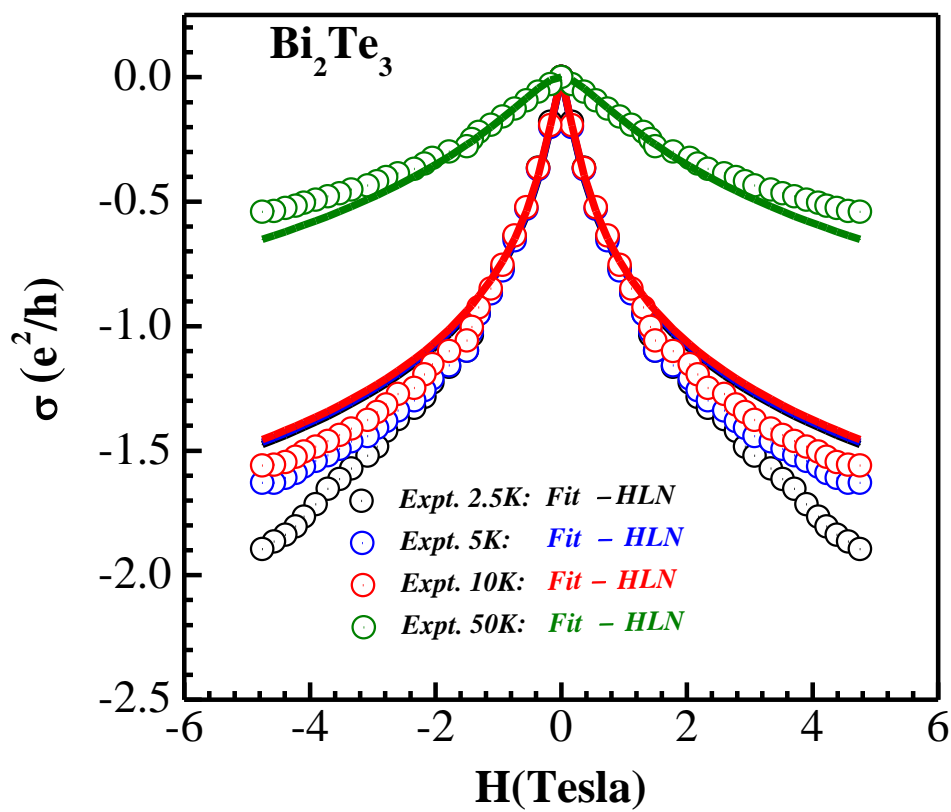

**Fig. 3**

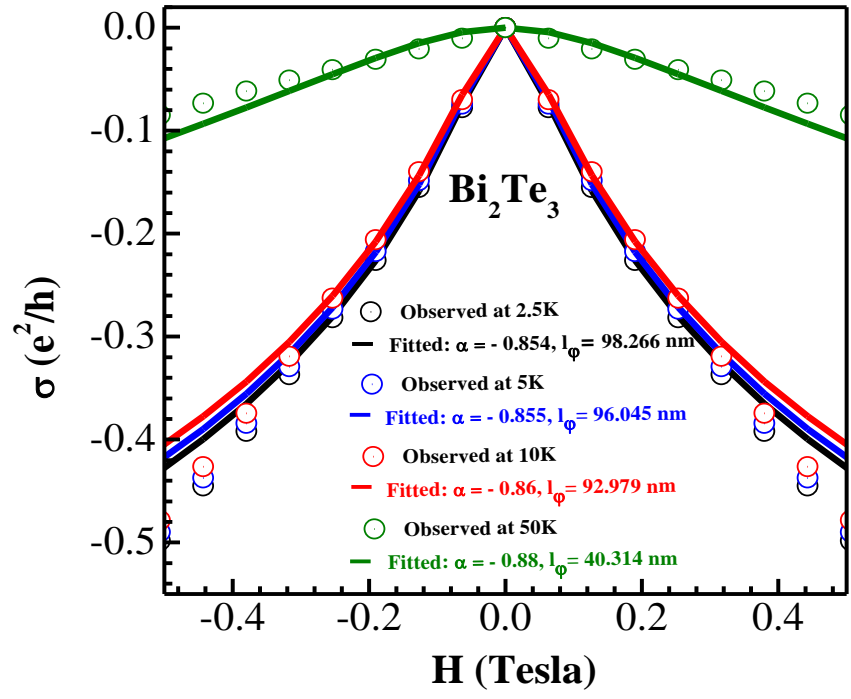